\documentclass[12pt, draftclsnofoot, onecolumn]{IEEEtran} 

    \usepackage[pdftex]{graphicx}
    \graphicspath{{.}}
    \DeclareGraphicsExtensions{.pdf,.png}

  \ifCLASSOPTIONonecolumn
    \newcommand{\defaultfigwidth}{0.8\linewidth}
  \else
    \newcommand{\defaultfigwidth}{0.99\linewidth}
  \fi

  \usepackage[cmex10]{amsmath}
  \usepackage{amsfonts}
  \newcommand{\beq}{\begin{IEEEeqnarray}{rCl}}
  \newcommand{\eeq}{\end{IEEEeqnarray}}

  \usepackage{multirow}

  \usepackage[nomargin,inline,draft]{fixme}
  \fxusetheme{color}

\begin{document}

  \title{Demonstration of multivariate photonics: blind dimensionality reduction with analog integrated photonics}

  \author{Alexander~N.~Tait\textsuperscript{*}, Philip~Y.~Ma, Thomas~Ferreira~de~Lima, Eric~C.~Blow, Matthew~P.~Chang, Mitchell~A.~Nahmias, Bhavin~J.~Shastri, and~Paul~R.~Prucnal
  \thanks{\textsuperscript{*}Corresponding author: atait@ieee.org}%
  \thanks{The authors are with the Department of Electrical Engineering, Princeton University, Princeton, NJ 08544, USA.
  A.N.T. is now with the National Institute of Standards and Technology, Boulder, CO 80305, USA.
  B.J.S. is with the Department of Physics, Engineering Physics \& Astronomy, Queen's University, Kingston, ON KL7 3N6, Canada.
  M.P.C. is now with Apple, Inc., Cupertino, CA 95014, USA.}}%


  \maketitle

\begin{abstract}
  \noindent
  Multi-antenna radio front-ends generate a multi-dimensional flood of information, most of which is partially redundant. Redundancy is eliminated by dimensionality reduction, but contemporary digital processing techniques face harsh fundamental tradeoffs when implementing this class of functions. These tradeoffs can be broken in the analog domain, in which the performance of optical technologies greatly exceeds that of electronic counterparts. Here, we present concepts, methods, and a first demonstration of multivariate photonics: a combination of integrated photonic hardware, analog dimensionality reduction, and blind algorithmic techniques. We experimentally demonstrate 2-channel, 1.0~GHz principal component analysis in a photonic weight bank using recently proposed algorithms for synthesizing the multivariate properties of signals to which the receiver is blind. Novel methods are introduced for controlling blindness conditions in a laboratory context. This work provides a foundation for further research in multivariate photonic information processing, which is poised to play a role in future generations of wireless technology.
\end{abstract}


\section{Introduction}
    The longstanding conception of ``the radio spectrum'' as a temporally and spatially homogeneous literal spectrum is eroding. Spatial inhomogeneities in the electromagnetic field, when exploited by multi-antenna or multi-input, multi-output (MIMO) systems, provide an orthogonal degree of freedom with which to share wireless resources~\cite{Biglieri:07}. Since 2009, multi-antenna communication techniques have been present in commercial standards for Wi-Fi (IEEE 802.11n~\cite{IEEE:09}), WiMAX (IEEE 802.16e), and LTE~\cite{Li:10}. Next generation wireless systems will use millimeter-wave frequencies ($f \approx$~60~GHz)~\cite{Kutty:16}, opportunistic access strategies~\cite{Akyildiz:08}, and/or massive MIMO ($N >$ 100 antennas)~\cite{Larsson:2014}. These prospects pose technical challenges for radio front-end hardware in terms of bandwidth, agility, and power efficiency. In addition, they each significantly increase demands for intelligent RF information processing. Taking into account the performance characteristics of electronic methods (both analog and digital), it becomes apparent that incremental improvements to the current state-of-the-art will not be enough to realize the full potential of spectrum access based on spatial discrimination.

    A large set of problems in multi-antenna front-ends fall within the category of dimensionality reduction. Signals at nearby antennas are typically highly redundant and must be transformed into a low-dimensional set of salient signals. The simplest dimension reducing operation is a linear projection. The simplest definitions of salience are the orthogonal directions of maximum covariance (the principal component (PC) vectors) and the directions of maximum statistical independence (the independent component (IC) vectors)~\cite{Hyvarinen2000}. Principal component analysis (PCA) yields decorrelated output signals, sorted in descending order of statistical relevance. The highest PCs can be discarded to reduce overall dimensionality while losing the minimum salient overall information. Independent component analysis (ICA) is equivalent to the problem of blind source separation (BSS), i.e., demixing signals that correspond to multiple source transmitters that have been mixed over a wireless channel. BSS is highly desired and widely applicable in many RF contexts because it can separate unknown mixtures of unknown signals using very weak \textit{a priori} assumptions about the signal characteristics. BSS can bring new capabilities to applications in spectrum sharing, electronic warfare, and multi-radio coexistence~\cite{Lee:05}.

    Multi-antenna dimensionality reduction presents two technical challenges. The first is finding the correct set of PC or IC vectors based on available observations, where ``correct'' means those that result in optimal output variance or independence, respectively. Algorithms for ICA have been studied extensively using digital signal processing (DSP) under the condition that all inputs are completely observable~\cite{Cardoso:93}. As discussed below, total observability is not always possible in relevant scenarios.

\begin{figure}[tb]
      \begin{center}
      \includegraphics[width=\defaultfigwidth]{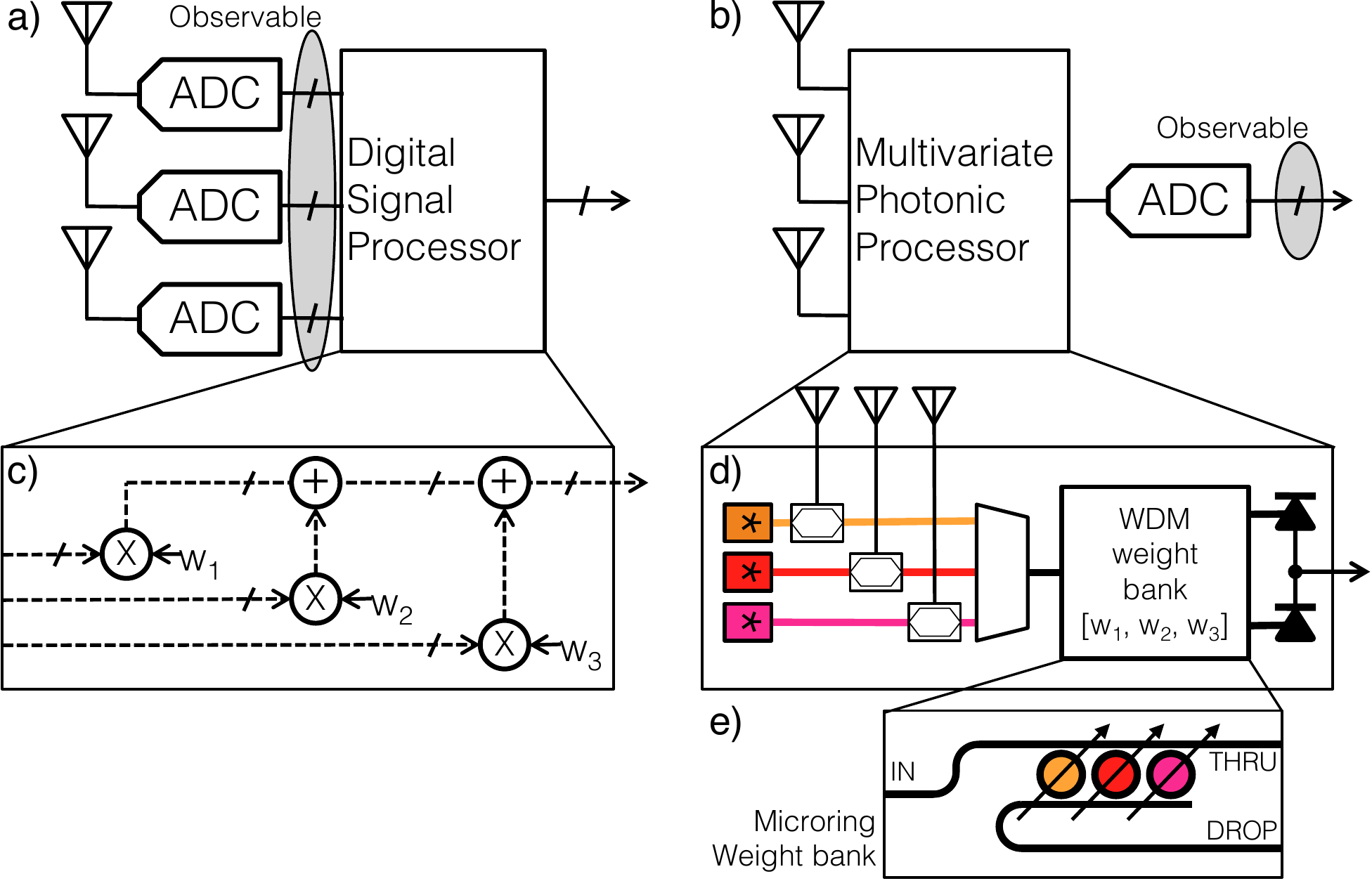}
      \caption[]{Comparison of radio front ends for dimensionality reduction. a) Electronic DSP based on ADC and MACs. b) Photonic weight bank based on modulation, WDM filtering, and photodetection. c, d) Internals of the dimension-reducing subsystem of the respective platforms. e) Microring (MRR) weight bank implementation of photonic weighted addition.}
      \label{fig:concept}
      \end{center}
    \end{figure}

    The second challenge is the scaling of energy needed to perform weighted addition operations with DSP hardware. A typical RF receiver, seen in Fig.~\ref{fig:concept}(a, c), consists of antennas, an analog domain, analog-to-digital converters (ADCs), and a digital signal processor. Every signal must be digitized before being processed, even though the majority of this information is non-salient and discarded in the digital domain~\cite{Gholam:11}. Both the multiply-accumulate (MAC) rate and the A/D conversion rate -- dominant power consumers~\cite{Walden:99,Murmann:15} -- scale in proportion to the product of number of dimensions and sampling frequency.

    The prohibitive performance tradeoff between dimensionality and power can be circumvented by reducing dimensionality in the analog domain. Multi-antenna beamformers implementing analog weighted addition have been explored~\cite{Venkateswaran:10,Gholam:11,Han:15a} and were reviewed in Ref.~\cite{Kutty:16}. These approaches -- typically known as hybrid analog-digital channel estimation -- have demonstrated significant reductions in power in dimension-reducing RF problems. The performance advantages of analog dimensionality reduction stem directly from \emph{not} digitizing all inputs. Non-salient information can be discarded in the analog domain, thus reducing the number of signal dimensions must be digitized. On the other hand, not digitizing corresponds to not observing all aspects of received signals. A comparison of observability in digital and analog is made between Fig.~\ref{fig:concept}(a, b). Constraints on observability give rise to a new challenge of how to solve estimation problems in the absence of conventionally necessary information~\cite{Ghauch:16}, a topic revisited in Sec.~\ref{sec:algorithm}.

    Analog, dimension-reducing front-ends are subject to the same hardware limits affecting the analog domain of all electronic front-ends. Analog RF electronics are inherently frequency dependent and so have severely limited performance in terms of bandwidth and agility (i.e. center frequency tunability)~\cite{Mongia:99}. These performance limits will be even more restrictive in next generation millimeter wave (mm-wave) systems with bands spanning 10s of GHz~\cite{Rappaport:11}. Contemporary multi-band wireless systems require a switched bank of RF circuits, each tailored to a single, narrow band~\cite{skyworks:11}. Switched bank architectures involve hardware that is greater in quantity and complexity and, furthermore, introduce additional signal degradation in the switch. These analog hardware burdens are multiplied by dimensionality in analog/hybrid channel estimation architectures since every channel needs its own switched bank.

    RF photonics have arisen to address limitations in the bandwidth and agility of analog RF electronics~\cite{Seeds:06,Marpaung:2013,Capmany:13}. In an RF photonic processor, an incoming signal is modulated onto an optical carrier wave, processed in some way, detected, and, only then, digitized. Modulation effectively upconverts an incoming signal to a 193~THz intermediate frequency (IF), which means that the signal in the optical domain has a minuscule \emph{fractional} bandwidth, even for 100~GHz bandwidth inputs. As a result, processing operations applied in the optical domain can be broadband and widely reconfigurable. Integrated RF photonics has offered improvements for programmable filtering~\cite{Perez:17,Zhuang:15,Liu:16a}, time delays~\cite{Liu:17,Burla:13,Khan:13,Cardenas:10}, transport~\cite{Marpaung:09,Chiles:17}, interference suppression~\cite{Liu:16,Chang:17}, and waveform generation~\cite{Khan:2010,Weiner:11}. These functions shown in RF photonics so far are relatively simple, single-input functions, much like the tasks performed in the analog domain of an RF electronic front-end.


    Multivariate photonics combines the advantages of analog photonic hardware and analog domain dimensionality reduction, employing statistical approaches to address the observability constraints that arise. Multivariate means multi-channel but is distinguished from multi-channel in that signals are treated as random variables instead of fully known waveforms. Adaptive dimensionality reduction based on statistical information also resembles learning approaches in photonic neural networks~\cite{Prucnal:17} and reservoir computers~\cite{Larger:12,Vandoorne:2014,Brunner:18}.

    In prior work, we proposed the concept of synthesizing multivariate signal features based on univariate observations of the analog photonic domain, which led to novel algorithms for performing blind PCA, ICA, and source separation~\cite{Tait:18ciss}. In this manuscript, we report the first demonstration of this approach by using a silicon photonic weight bank to ascertain the PC dimensions of and then decompose unobservable 1.0~GHz signals. We introduce novel methods -- such as how to control observability constraints in a laboratory setting -- that will support further experimental research in multivariate photonics.

    Multi-antenna front-ends implemented by conventional electronics and multivariate photonics are juxtaposed in Fig.~\ref{fig:concept}. In the photonic front-end in Fig.~\ref{fig:concept}(b), dimensionality reduction occurs in the analog domain, after which, only one signal is digitized. The result is a fundamental alteration of power scaling trends:
    \beq
      P_{{DSP}} &=& Nf (E_{{ADC}} + E_{{MAC}}) \\
      P_{{photonic}} &=& N P_{{laser}} + f E_{{ADC}}
    \eeq
    where $N$ is dimensionality, $f$ is bandwidth, $E_{{ADC}}$ is the ADC conversion energy, $E_{{MAC}}$ is MAC energy, and $P_{{laser}}$ is the power needed for each laser pump. We note that ADCs can be time-domain multiplexed to trade off hardware with bandwidth~\cite{Yang:16}, yet these approaches exhibit the same DSP power relations. This difference in power scaling is significant because the second equation has no dominant $Nf$ term. Power to modulate does scale with $Nf$, but it is negligible (1~fJ~$\times f$~\cite{Timurdogan:14}) in comparison to ADC powers (1~pJ~$\times f$~\cite{Murmann:15}). This means that multivariate photonics could radically reduce power in high-speed, high-dimensionality regimes.

    Wavelength-division multiplexed (WDM) weighted addition has been proposed as an efficient implementation of multi-input information processing in the analog photonic domain~\cite{Tait:14}. Weighted addition is naturally implemented using power-modulated, mutually incoherent optical carriers so that fan-in of carriers does not cause coherent interference~\cite{Goodman:1985}. WDM carriers can divide the THz-wide transmission window of a single waveguide among many GHz-bandwidth signals, but the hardware requirements of fiber optics pose a practical limit to channel count.

    Silicon photonic integration brings to optics unprecedented opportunities in both large-scale system integration and large-volume economies of scale~\cite{Hochberg:13,Thomson:16}. These opportunities impact unconventional signal processing approaches to the extent that the approach can be implemented with conventional silicon photonic devices. WDM weighted addition is implemented in silicon photonics using microring (MRR) weight banks~\cite{Tait:16scale}. An MRR weight bank can weight WDM channels independently over a continuous and balanced (--1 to +1) range with a currently demonstrated accuracy of 5.1 bits~\cite{Tait:16multi,Tait:18fb}.

    Photonic techniques for multivariate processing in prior work. Photonic ICA was first developed in holographic materials~\cite{Anderson:04} that are not readily integrable. Principal component analysis was explored in fiber~\cite{Tait:15,FerreiradeLima:16} and MRR weight banks~\cite{Tait:17cleo}. To ascertain the original inputs, these latter three works scanned through identity projections of the form $\left[1, 0, 0, \ldots\right]$. This strategy is unrealistic because it assumes that inputs are synchronized and repeating with a period known to the receiver performing PCA. The approach in this work does not make any of these assumptions; it is blind to the input waveforms. In Sec.~\ref{sec:algorithm}, we describe the information that can be extracted from multi-dimensional signals whose waveforms are unobservable. In Sec.~\ref{sec:blindEnvironment}, we describe how blindness can be ensured and evaluated in an experimental setting.

    \begin{figure}[tb]
      \begin{center}
      \includegraphics[width=\defaultfigwidth]{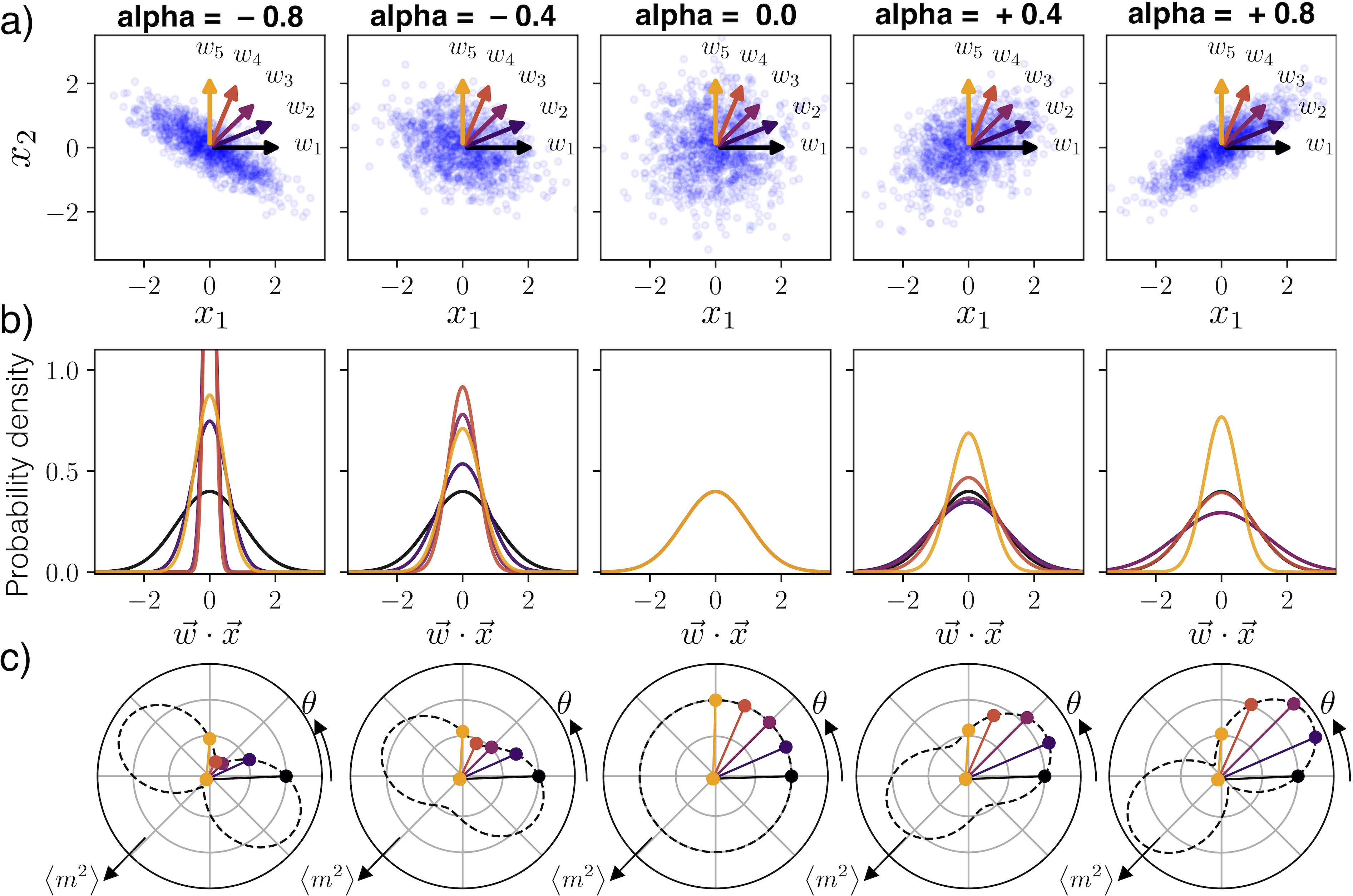}
      \caption[]{Visual interpretation of projected bivariate statistics over 5 values of cross-correlation, $\alpha$. a) Blue points: synchronously sampled 2-channel signals $\vec{x}(t)$, unobservable to an analog front-end; colored arrows: projection vectors, $\vec{w}_i$. b) Histograms of each projected signal $\vec{w}_i \cdot \vec{x}(t)$. Curve color corresponds to the projection vectors in (a). c) Polar plots of projection variance (i.e. second moment) vs. projection angle, $\theta$. Line segments: measured points with color corresponding to above projection vectors, $\vec{w}_i$; dashed curve: full model fit to points. Its major and minor axes are the first and second principal component vectors, respectively.}
      \label{fig:epiTheory}
      \end{center}
    \end{figure}

\section{Using projected moment information} \label{sec:algorithm}
    Analog-domain dimensionality reduction faces a key barrier to finding the weight vectors of interest based on a reduced set of information. As a direct consequence of discarding information prior to digitization, it is not possible to take synchronous samples of multiple signal dimensions. Consequently, dimension-reducing analog front-ends cannot recover signal covariation and cannot employ on traditional techniques for PCA and ICA that rely on having simultaneous samples of all input signals~\cite{Lee:05}. We refer to this property as an observability constraint.

    Recent strides have been made to address this constraint in hybrid channel estimation. They have been based on different \textit{a priori} assumptions, such as cooperative scenarios with bidirectional communication~\cite{Alkhateeb:14,Ghauch:16,Venugopal:17,Zanjani:17,Dahl:04}. Many of these approaches resemble modern WiMAX protocols that use training signals sent during an estimation phase (exception by Dahl et al.~\cite{Dahl:04}). Another common assumption is channel sparsity (exception by Zanjani et al.~\cite{Zanjani:17}). Iltis et al.~\cite{Iltis:06} studied a non-cooperative scenario using a game theoretic approach, which must assume that all transceivers abide by a common protocol.
    In~\cite{Tait:18ciss}, we proposed an algorithm for performing BSS when only one reduced-dimension output can be observed at a given time. The only \textit{a priori} assumption used is that the sources to be separated are statistically independent: a weak and generally true assumption. To the authors' knowledge, this type of fully blind technique has not been proposed in the research on electronic hybrid channel estimation, even though it could also be compatible with electronic hardware.


    Instead of attempting to recover synchronized multi-channel waveforms, the approach acts based on synthesizing multiple statistical measurements of one-dimensional projections. Figure~\ref{fig:epiTheory} illustrates relationships between samples, projections, and statistical measurements. Two-dimensional simultaneous samples are shown in Fig.~\ref{fig:epiTheory}(a) as blue points -- these are not observable with a multi-channel analog/hybrid front-end. Columns correspond to different cross-correlation values, alpha ($\alpha$). More correlated signals result in point clouds that are more stretched in the 45$^\circ$ direction, while more anti-correlated (negative alpha) signals are more stretched at --45$^\circ$. Projections of the inputs result in a one-dimensional output representing the dot product of weights: $m(t) = \vec{w} \cdot \vec{x}(t)$, where $\vec{x}$ are the received signals that have been mixed over the air and $m$ is their linear projection along the $\vec{w}$ vector. Five example weight vectors with equal magnitude and different angles are overlaid in Fig.~\ref{fig:epiTheory}(a).

    The key measurement involved in this approach is of statistical moments of $m$, which describe the width of a one-dimensional histogram. The histograms of $m$ are shown in Fig.~\ref{fig:epiTheory}(b) -- these are observable. Colors correspond to the projection vectors in Fig.~\ref{fig:epiTheory}(a). The widths of these histograms depend on the angle of the weight vectors and the cross-correlation of the inputs. If the weight vectors are known, the cross-correlation can be found. Mathematically, a moment is the expectation value or time-average of a random variable raised to a power. The $b$\textsuperscript{th} moment of the output is denoted as $\left<m^b\right>_t$, where $\left<\cdot\right>_t$ is a time average.

    The use of moment measurements has three main advantages in terms of relaxing sampling requirements. Firstly, it is not necessary to sample above the Nyquist rate needed to reconstruct a waveform. The subsampling theory states that an unbiased, random sampling of a dataset has the same histogram as that original dataset, provided some sample size conditions on which we won't elaborate here. In Fig.~\ref{fig:fullSetup}(b-d), we show an experimental implicate of that theorem -- reproducing the histogram of a Nyquist-sampled waveform using samples taken at 10$^{-5}$ of the Nyquist rate. Secondly, all signals have moments and yield to statistical analysis. Unlike waveform analysis that is format-specific, statistical analysis naturally generalizes to the heterogeneous formats and usage types in the diverse wireless space.

    Finally, moments are time-invariant when the channel mixing is stationary. This means that multiple measurements of time-averages can be made relatively far apart in time and still give information about a consistent cross-correlation structure. In an RF context, stationary timescales are related to environmental and channel fluctuations, which are on the order of seconds. While there is no theoretical minimum sample rate needed to measure a moment, in practice, faster sampling translates to an ability to make more moment measurements per stationary interval.

    The key realization of~\cite{Tait:18ciss} is that multi-dimensional correlation and independence can be inferred and exploited based only on measurements of projection moments.
    The moments of the projections have a simple relationship to the projection angle. Constraining $\vec{w}$ to unit normal vectors, they can be parameterized by an angle, $\theta$, such that $\vec{w} = \left[ \cos\theta, \sin\theta \right]$. Regardless of the characteristics of $\vec{x}$, the relationship of moment to angle follows the model:
    \beq
      \left<m^2\right> (\theta) &=& q_1 + q_2 \cos\left[2(\theta - \theta_0)\right] \label{eq:twoPetal}
    \eeq
    where $q_1$, $q_2$, and $\theta_0$ are stationary parameters. Figure~\ref{fig:epiTheory}(c) shows polar plots of this model of of $\left<m^2\right>$ vs. $\theta$. Line segments represent the moments as measured at the angles of the weight vectors in Fig.~\ref{fig:epiTheory}(a) with corresponding colors. Black dashed curves show the full models described by Eq.~\ref{eq:twoPetal} that fit these points.

    There is a clear visual correspondence between the black dashed curves in Fig~\ref{fig:epiTheory}(c) and the shape of point cloud distributions in Fig.~\ref{fig:epiTheory}(a). Complete knowledge of $q_1$ and $\theta_0$ is equivalent to complete knowledge of all 2\textsuperscript{nd}-order covariant statistics of the input distribution. The magnitude of the first principal component is $q_1 + q_2$, and that of the second is $q_1 - q_2$. The first PC vector angle is $\theta_0$, and the second is $\theta_0 + \pi/2$. In higher dimensions, a similar model with more $q$ and $\theta$ parameters still holds.

    In the algorithm to perform PCA based on moment observations, the parameters are found by fitting this model to multiple measurements of moments at different projection angles. A photonic weight bank is reconfigured to project the inputs along a given angle, the 2\textsuperscript{nd} moment of the summed output is recorded, then the weight bank projects along the next angle and so on. PCA alone is generally not particularly useful in terms of RF signal separation (exception in~\cite{Bhatti:12}), but it is closely related to and the first step of BSS. We refer the reader to~\cite{Tait:18ciss} for more detail on the fitting algorithm, as well as the extension to 4\textsuperscript{th}-order moments, higher dimensions, and blind source separation. For the remainder of this paper, we will focus just on blind PCA in two dimensions and its demonstration with a silicon photonic weight bank.

\section{Methods}

  \subsection{Correlated WDM signal generation} \label{sec:siggen}
    PCA is a decorrelation technique, so, to evaluate PCA, we must start with signals that have varying degrees of correlation. We construct a signal generator that produces multi-wavelength signals whose cross-correlation can be programmed at will, following a method introduced in Ref.~\cite{Tait:15}.

\begin{figure}[tb]
      \begin{center}
      \includegraphics[width=\defaultfigwidth]{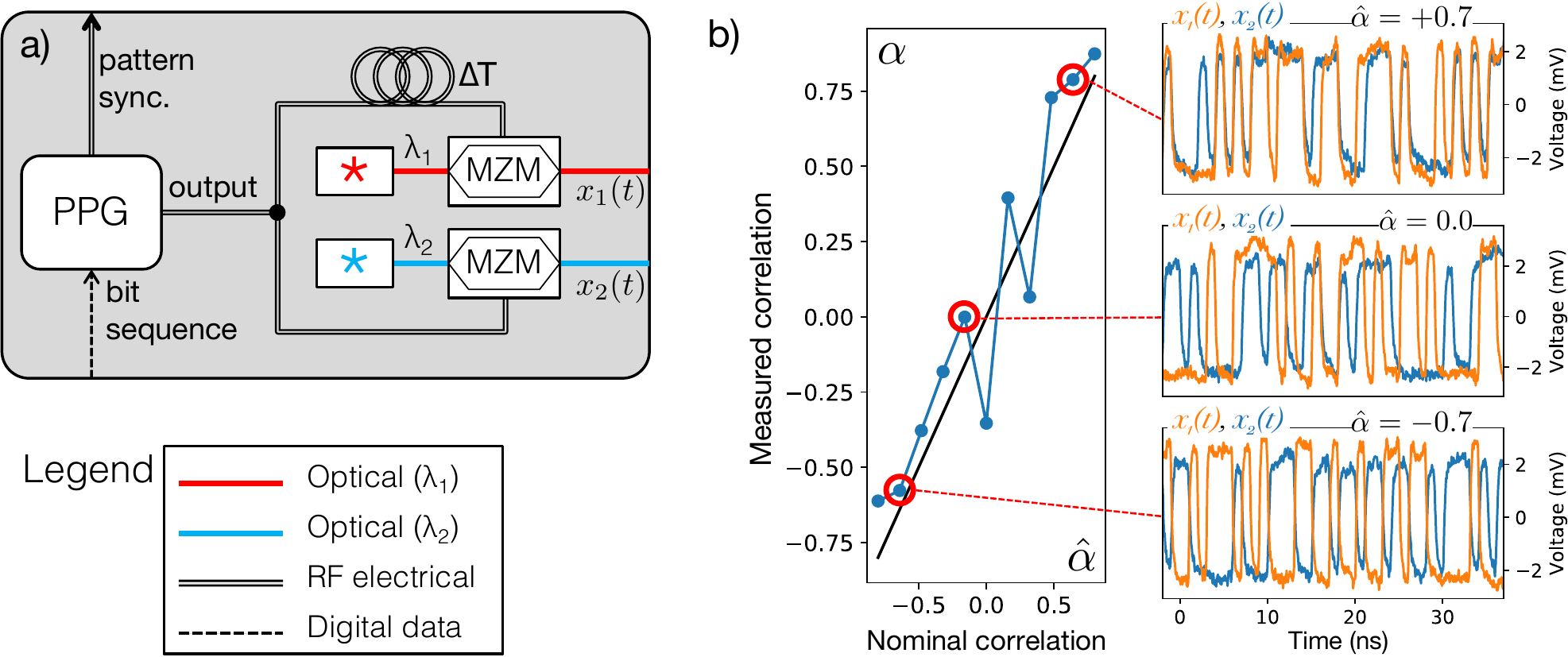}
      \caption[]{Signal generator subsystem. a) A single pulse pattern generator (PPG) modulates two optical carriers with RF time skew of $\Delta T$. The programmed bit pattern is an autocorrelated sequence, derived from a nominal cross-correlation ($\hat{\alpha}$) and a knowledge of $\Delta T$. b) Measured cross-correlations where $\hat{\alpha}$ is both the nominal cross-correlation and auto-correlation of the programmed bit pattern. Black line indicates equality between measured and nominal. Waveform pairs after optical modulation illustrate the probability of corresponding bits being the same.}
      \label{fig:sigGen}
      \end{center}
    \end{figure}

    The input generation subsystem (Fig.~\ref{fig:sigGen}a) consists of a single pulse pattern generator (PPG), modulators, and channel-dependent time delays spaced by $\Delta T$. The time skew has the effect of transforming temporal auto-correlation of the original PPG signal to instantaneous cross-correlation between the outputs. The experimental signals used (binary) bear little relation to real radio signals (analog, modulated); however, they can still be treated as analog waveforms that can be meaningfully weighted and summed. Their time-averaged correlation is a continuous value.  The reason for using binary signals is that their statistical properties are easily controlled by programming particular bit patterns to a PPG.

    The degree of cross-correlation is controlled by programming appropriate bit patterns. First, the time skew, $\Delta T$, is measured, in this case, 13.54 ns. The PPG bit rate, $B$, is then set to 1.9956~GHz so that the time skew is an integer multiple of the bit period: $B \Delta T = 27.00$. The programmed bit sequence is initialized with a 27-bit pseudo-random bit sequence (PRBS). The 28$^{th}$ bit has some probability of being the same as the 1$^{st}$ bit, and so on. This probability is derived from to the desired correlation value:
    \beq
      P(b_{i} = 1) = 0.5 + \hat{\alpha} (b_{i-27} - 0.5)
    \eeq
    where $\hat{\alpha}$ is the desired (i.e. nominal) correlation, $P$ means probability from 0 to 1, and $b_i$ is the value of bit $i$.
    The resulting sequence is a bit delayed, first-order Markov process. To extend this generation technique to more than two channels, a multi-order Markov process would be used, as studied in~\cite{FerreiradeLima:16}. Unlike prior methodology~\cite{Tait:15}, this generation procedure does not require $\Delta T$ to be set precisely.

    We evaluate the capabilities of this signal generator over a range of $\hat\alpha$'s from --0.8 to +0.8. The real correlation is $\alpha = \frac{\left<x_1 \cdot x_2\right>}{\left<x_1^2\right>\left<x_2^2\right>}$, where $x_1$ and $x_2$ are measured in an oscilloscope. Figure~\ref{fig:sigGen}(b) indicates that the correlation between multiwavelength inputs is controlled by $\hat\alpha$, even though $\hat\alpha$ parameterizes a digital stochastic process while $\alpha$ is an analog concept.

    The cross-correlation was observed to match the value of $\alpha$ within an error of 0.20 with more error for correlations around zero. This value is higher than expected due to impedance mismatches causing spurious correlations, discussed further in Sec.~\ref{sec:discussion}. $\alpha$ can range from $-1$ (perfect anticorrelation, all bits separated by 27 are different) to $0$ (uncorrelated, pseudo-random bit sequence) to $+1$ (perfect correlation, all bits separated by 27 are the same).

    \begin{figure}[tb]
      \begin{center}
      \includegraphics[width=\defaultfigwidth]{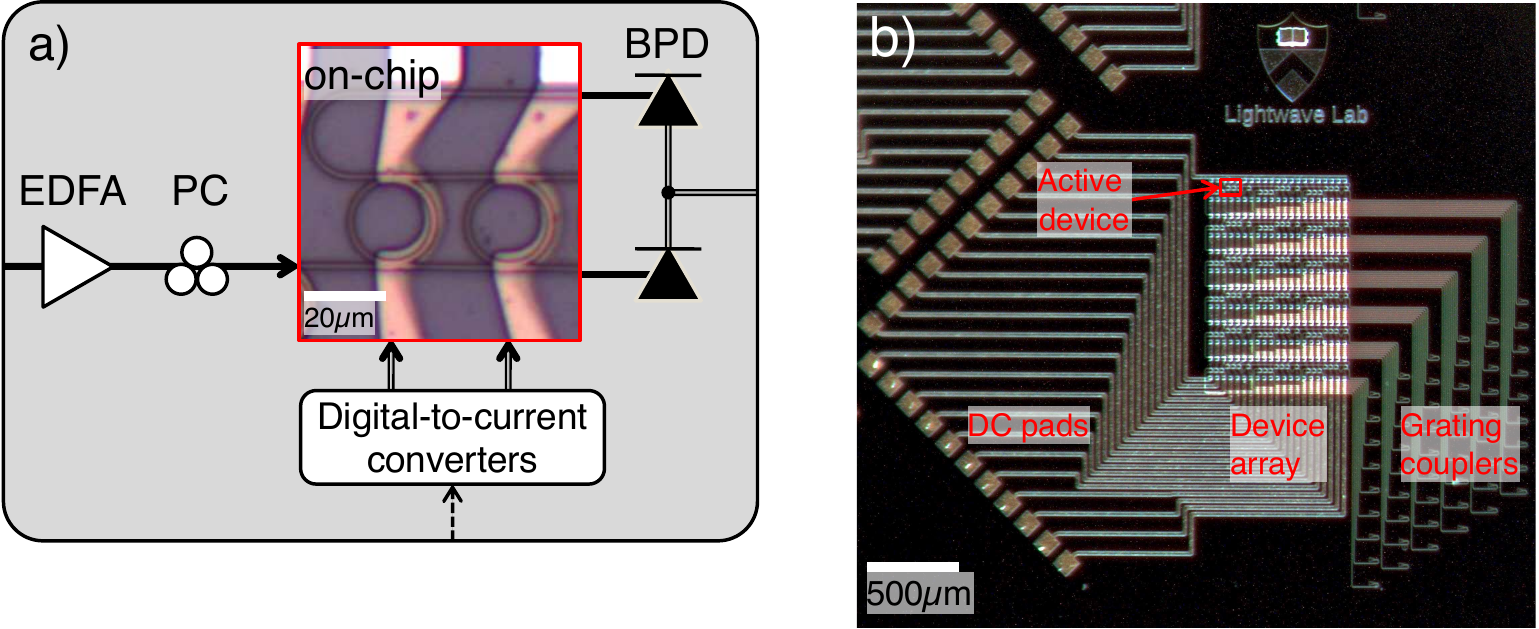}
      \caption[]{a) WDM weight bank subsystem. A WDM input is coupled to a silicon photonic chip to a microring (MRR) weight bank. EDFA: erbium doped fiber amplifier; PC: polarization controller. Its complementary outputs are detected off chip to produce an electrical weighted sum output. BPD: balanced photodetector. The weight values are configured by heating resonators individually via metal filaments. b) Overview of the sample containing a large test array of MRR weight bank devices reusing electrical ports and grating couplers. In this array, there are 22 electrical ports, 6 sets of 9 GCs each, and 126 MRRs in various circuit configurations. Only one bank is actively addressed at a time.}
      \label{fig:wbDevice}
      \end{center}
    \end{figure}

  \subsection{Silicon microring weight banks}
    The microring weight bank pictured in Fig.~\ref{fig:wbDevice} is fabricated on a silicon-on-insulator die with a 220nm device layer. 500nm wide waveguides are patterned by EBeam lithography and fully etched to the buried oxide~\cite{Bojko:11}. After the waveguide etch, a 3$\mu$m oxide passivation layer and metals are deposited. The sample is mounted on a temperature controlled alignment stage and coupled to fiber through focusing sub-wavelength grating couplers~\cite{Wang:14opex}.

    In the weight bank, the MRRs have a short 2$\mu$m straight coupling region with arcs of slightly different radii near 10$\mu$m. They have FSRs of 9.1nm and quality factors of 22,000. The radii are different by a minuscule 16 nm such that the as-fabricated inter-channel resonance offset is nominally 2.3nm. The actual offset is affected by fabrication variation.

    Thermal tuning filaments are formed of a Ti/Au layer and are routed to pads in an Al routing layer~\cite{antFab}. Electrical measurements yield sheet resistivities of 0.17~$\Omega/\square$ (routing) and 6.7~$\Omega/\square$ (filament). Contact resistance is 3.2~$\Omega$, and routing-filament via resistance is 42~$\Omega$. Based on these values, we estimate an internal filament resistance of 52~$\Omega$. Using this estimate and measurements of resonance shift vs. tuning power, we find that internal tuning efficiency is 0.22nm/mW, in other words, a FSR power of $P_{\mbox{FSR}}$=42 mW.

    The convergence of the PCA algorithm depends on the ability to command multiple weights accurately and simultaneously. Microring weights are controlled using the feedforward approach developed in Ref.~\cite{Tait:16multi}. First, an optical spectrum analyzer (OSA) is used to independently track MRR resonances onto their corresponding WDM signal wavelengths. Thermal cross-talk is measured around this operating point and, later, counteracted by the control algorithm. Finally, each MRR is detuned through its resonance edge while real weight is measured externally to obtain a one-to-one monotonic function of weight vs. detuning. During the training phase, multiple weights are measured simultaneously by decomposing calibration signals chosen such that they are uncorrelated. In future work, the feedback weight control shown in~\cite{Tait:18fb} could be used to accomplish the same goal with more resilience and accuracy.

    \begin{figure}[tb]
      \begin{center}
      \includegraphics[width=\defaultfigwidth]{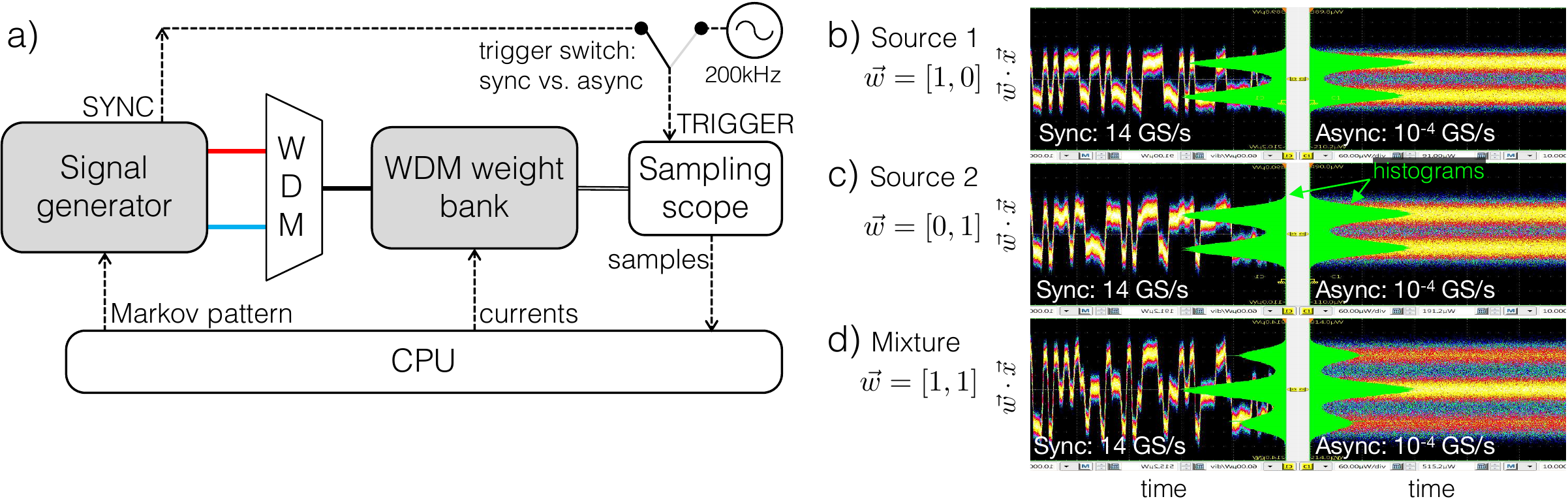}
      \caption[]{a) Full experimental setup showing signal generation, WDM weighting, and measurement. Each subsystem is controlled by computer (CPU). The scope is triggered by either the pattern sync of the pattern generator (sync state) or a free-running clock (async state). b) Scope views over both synchronization states and three projection vectors. In asynchronous mode (right column), the effective sample rate is deeply sub-Nyquist, so waveform and timing information is lost. Nevertheless, signal histograms (green) are identical to those of the super-Nyquist waveforms, provided a sufficient number of samples are taken.
      }
      \label{fig:fullSetup}
      \end{center}
    \end{figure}

  \begin{figure}[tb]
    \begin{center}
    \includegraphics[width=\defaultfigwidth]{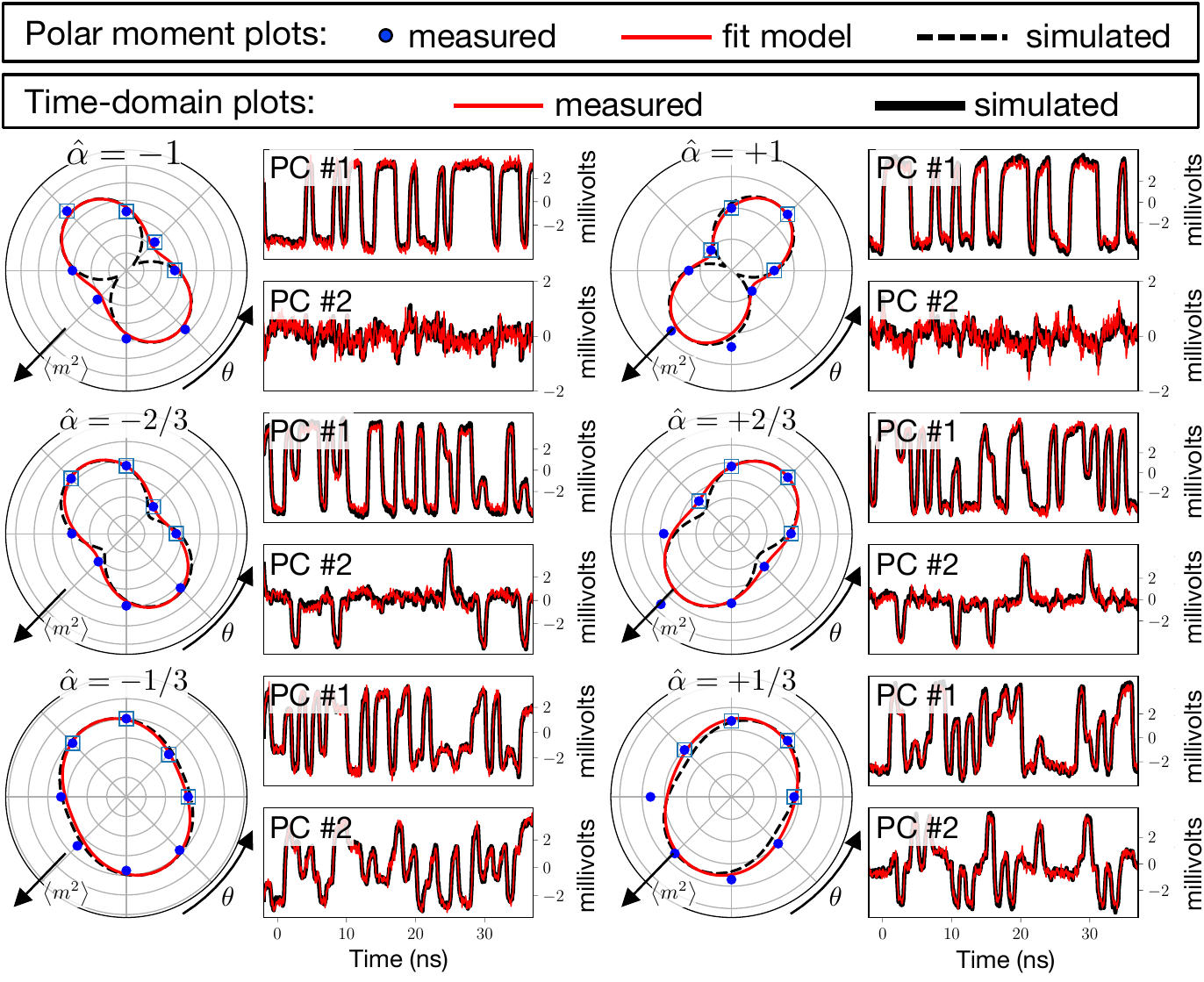}
    \caption[]{Experimental data over a range of six partial correlation parameters, $\alpha$, one in each panel. Polar plots depict variance vs. angle that is measured (blue dots), fit (red curve), and predicted (black dashed curve). The two principal component (PC) waveforms are plotted on the right side of each panel. Those found by the proposed photonic procedure (red) reproduce those predicted based on complete knowledge of the inputs (black).}
    \label{fig:pcaData}
    \end{center}
  \end{figure}

  \subsection{Controlling observability through synchronization} \label{sec:blindEnvironment}
    To evaluate the accuracy of blind multivariate algorithms in the laboratory, Nyquist-sampled waveforms must be observed and compared against expectation. At the same time, the waveforms must be obscured while the algorithm is running in order to produce a situation corresponding to a real-life (a.k.a. field) scenario. In a non-cooperative field scenario, waveforms are non-periodic, and the source and receiver share no synchronous trigger. In other words, time information is lost, but amplitude information is present. If waveforms are not obscured, the periodic laboratory artifacts can be exploited to demonstrate laboratory procedures that then cannot be applied to the field~\cite{Tait:15,FerreiradeLima:16,Tait:17cleo}.

    We construct a setup that can easily switch between a state corresponding to the field (waveforms obscured) and a state needed to evaluate performance (waveforms observable). Timing information is obscured on the hardware level by disrupting the synchronization between the PPG and sampling oscilloscope, a Tektronix DSA8300. The trigger switching setup is shown in the experimental diagram in Fig.~\ref{fig:fullSetup}a. In the synchronous state, the PPG pattern sync output triggers the scope, allowing it to emulate a super-Nyquist real-time scope (here 14~GS/s). In the asynchronous state, a free running clock triggers the scope, creating a situation equivalent to deeply sub-Nyquist real-time scope: here, $2\times 10^{-4}$~GS/s.

    This method requires an understanding of how sampling scopes operate. A sampling scope has a very low real-time sampling rate, yet it can take advantage of periodicity in laboratory-generated signals to achieve a high synthetic sampling rate. In each period, a sample is taken with a slightly different delay from the beginning of the period. After many periods, these samples are reordered to construct a complete, high-sample-rate picture of the periodic waveform. This process depends on the ability of the sampling scope to synchronize to the instrument generating the periodic signal. Disconnecting the trigger destroys any capability to recover timing information, thereby creating a situation closely resembling a field scenario.

    Fig.~\ref{fig:fullSetup}(b-d) illustrates the difference in observability caused by the trigger switching. In moving from synchronous to asynchronous sampling, waveform information is lost, but voltage histograms (green) are maintained. This illustrates, firstly, that deeply sub-Nyquist sampling successfully obscures timing information, and, secondly, that it still provides the statistical information required to exploit projected moment measurements. Switching between synchronization states satisfies the conflicting goals of emulating a realistic field scenario and evaluating the efficacy of the algorithm. This technique for obscuring waveforms with a simple modification to a typical laboratory setup is here used to evaluate a blind PCA algorithm, but it could broadly extend to evaluate a wide variety of blind RF algorithms, including blind source separation.

\section{Results}
  During the photonic PCA procedure, the trigger is in the asynchronous state. The algorithm described in Sec.~\ref{sec:algorithm} -- and in more detail in~\cite{Tait:18ciss} -- is executed. The weight vectors applied by the MRR weight bank are tuned through a range of angles while their magnitude is held constant. At each point, the weighted sum is sampled at 200~kS/s. The moment of these samples is calculated and plotted in Fig.~\ref{fig:pcaData} as blue dots. These measurements are fit with the model of Eq.~\eqref{eq:twoPetal} (red polar curves) using the Gauss-Legendre pseudo-inverse method described in~\cite{Tait:18ciss}. From this fit, the PC vectors are ascertained. This signal generation and PCA procedure are repeated over a range of three negative and three positive covariance parameters.

  After the PCA algorithm is performed, accuracy is evaluated. The trigger is set to the synchronous state, and input waveforms are recorded. Using this waveform information, the CPU calculates the expected PCs and PC vectors using a traditional singular value decomposition algorithm. Based on the PC vectors, the expected model of Eq.~\ref{eq:twoPetal} is generated and plotted as black dashed lines in the polar plots. The PC signals are shown as the black waveforms in the time-domain panels. Next, the weight bank is tuned to the PC vectors \#1 and \#2 as determined during the asynchronous procedure. The actual output waveforms are recorded, shown as the red curves.

  The actual and expected outputs are compared based on the normalized root mean squared (RMS) error of these waveforms in Table.~\ref{tab:results}. The first column, $\hat{\alpha}$, is the nominal correlation used to generate signals as described in Sec.~\ref{sec:siggen}. The measured correlation of inputs, $\alpha$, is in the next column. The ``signal'' column shows the RMS voltage of the expected PC waveforms as determined by the conventional CPU algorithm. The RMS of the first PC is higher with larger absolute value of correlation because the joint distribution is more stretched, as is seen in Fig.~\ref{fig:concept}a.

  The noise amplitude here is considered to be the signal components above 2.5~GHz that are not repeatable between multiple recordings of the inputs. It would be possible to instead define noise as departures from a clean binary signal, but we do not use this approach because the input signals are treated as analog. The repeatable ripples and deviations away from a clean binary signal are considered part of the actual input. The noise cutoff frequency is chosen to roughly separate these repeatable components from the non-repeatable components.

\begin{figure}[tb]
    \begin{center}
    \includegraphics[width=.6\linewidth]{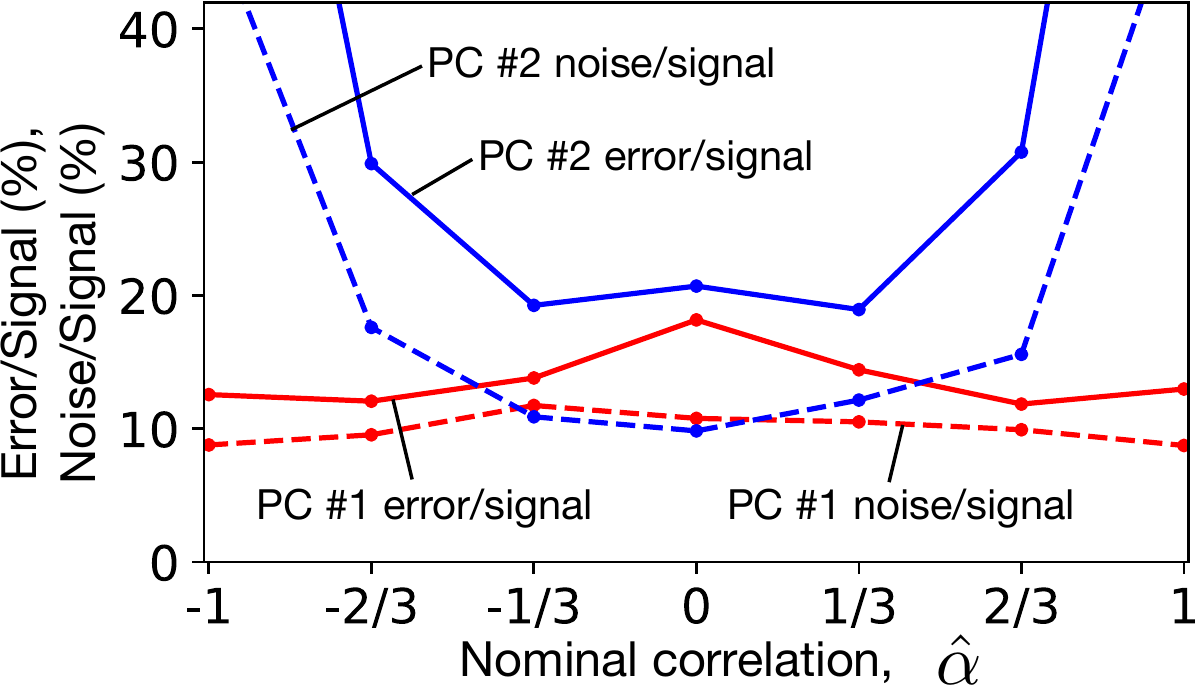}
    \caption[]{Plot of values in Table~\ref{tab:results}. Solid lines are relative error, and dashed lines are relative noise. Red lines are for PC \#1 and blue for PC \#2. Noise is a significant component of error, particularly visible in PC \#2.}
    \label{fig:metaData}
    \end{center}
  \end{figure}

  Error is the RMS of the difference between the measured (red) and expected (black) PCs. Error and noise, normalized by the signal amplitude, are plotted in Fig.~\ref{fig:metaData} to visualize their trends. It can be seen that it is easier to find the first component when absolute correlation is high. Since the PC \#1 signal is larger, signal-to-noise ratio is also larger. At these points, the PC \#2 signal has a vanishing amplitude, resulting in vanishing signal-to-noise ratio and correspondingly more error. At all points, noise can account for a majority of the total error amplitude. As such, a system with reduced noise would have less error.

  Nominal correlation of 0 is a special case. PC \#2 has its largest amplitude resulting in a noise minimum; however, the PCA problem is close to degenerate, so the error of PC \#2 does not have a minimum. In theory, the error at $\hat\alpha=0$ should approach 100\%, but, as seen in Table~\ref{tab:results}, the actual correlation is not quite zero, so the PCA algorithm does have a well defined solution. This spurious correlation could be due to reflections in the RF circuit off of modulator 1 that find their way to modulator 2. Neglecting the degenerate $\hat\alpha=0$ trial, the error of the first PC is always less than 14\%.

  \begin{table*}
    \caption{RMS values in millivolts of signal, noise, and error for both PCs over the range of correlation values evaluated. Percent error is ratio of error to signal. The noise-error ratio indicates that measurement noise contributed significantly to error.} \label{tab:results}
    \small \centering
    \begin{tabular}{|c|c|c|c|c|c|c|c|}
      \hline
      $\hat{\alpha}$ & $\alpha$ & PC\# & Signal (mV) & Noise (mV) & Error (mV) & Error (\%) & Noise/Error (\%) \\
      \hline
      \hline

      \multirow{2}{*}{$-1$} & \multirow{2}{*}{--0.97} &  1 &  3.06 &  0.27 &  0.38 &  13 &  70 \\
      \cline{3-8}
      & &  2 &  0.35 &  0.17 &  0.31 &  90 &  54 \\
      \hline

      \multirow{2}{*}{$-\frac{2}{3}$} & \multirow{2}{*}{--0.76} &  1 &  3.41 &  0.33 &  0.41 &  12 &  79 \\
      \cline{3-8}
      & &  2 &  1.22 &  0.21 &  0.36 &  30 &  59 \\
      \hline

      \multirow{2}{*}{$-\frac{1}{3}$} & \multirow{2}{*}{--0.20} &  1 &  2.37 &  0.28 &  0.33 &  14 &  85 \\
      \cline{3-8}
      & &  2 &  1.80 &  0.20 &  0.35 &  19 &  57 \\
      \hline

      \multirow{2}{*}{$0$} &  \multirow{2}{*}{0.12} &  1 &  2.42 &  0.26 &  0.44 &  18 &  59 \\
      \cline{3-8}
      & &  2 &  1.98 &  0.19 &  0.41 &  21 &  48 \\
      \hline

      \multirow{2}{*}{$+\frac{1}{3}$} &  \multirow{2}{*}{0.31} &  1 &  2.52 &  0.27 &  0.36 &  14 &  73 \\
      \cline{3-8}
      & &  2 &  1.76 &  0.21 &  0.33 &  19 &  64 \\
      \hline

      \multirow{2}{*}{$+\frac{2}{3}$} &  \multirow{2}{*}{0.67} &  1 &  3.27 &  0.32 &  0.39 &  12 &  84 \\
      \cline{3-8}
      & &  2 &  1.42 &  0.22 &  0.44 &  31 &  51 \\
      \hline

      \multirow{2}{*}{$+1$} &  \multirow{2}{*}{0.98} &  1 &  3.66 &  0.32 &  0.48 &  13 &  67 \\
      \cline{3-8}
      & &  2 &  0.34 &  0.17 &  0.34 &  101 &  51 \\
      \hline
    \end{tabular}
  \end{table*}

\section{Discussion} \label{sec:discussion}
  Multivariate photonics is a new and uncharted direction for high-bandwidth, nontrivial information processing with photonic systems. Photonics presents potentially disruptive opportunities for future wireless techniques, but it has yet to prove its viability and motivation as a commercial technology. Further research in this field depends on the answers to several fundamental research questions and corresponding experiments.

  \subsection{Dynamic range}
    An important application of BSS is blind interference suppression, wherein the interferer can be many orders of magnitude more powerful that the signal of interest, thus requiring a high dynamic range to resolve the weak signal. Analog photonic technologies possessing high dynamic range have been shown to excel at interference suppression when the interferer is out-of-band~\cite{Capmany:06,Aryanfar:16,Perez:17} and when the interferer is in-band and the interfering signal is is known~\cite{Chang:17,Zhou:14,Sun:17}. For example, Ref.~\cite{Chang:13} demonstrated cancellation of 60~dB over a 50 MHz band, limited by electronic impedance mismatches in the setup. Since it is based on the same hardware, blind photonic interference cancellation can be expected to dynamic ranges similar to known-interference cancellation systems. Besides interference depth, analog photonic links have tradeoffs in dynamic range, power, and bandwidth that differ from corresponding electronic tradeoffs~\cite{Marpaung:09}. Further work comparing these tradeoffs to DSP is called for.

    The ability of photonic BSS to achieve this performance also rests on the algorithm's ability to converge. Since this aspect is not present in known-interference problems found in the literature, it will require new simulation and experimental research. It is possible that there is a limiting signal-to-noise/interference-ratio (SNIR) below which the algorithm will fail to converge entirely. Studying this limit is necessary to understand which scenarios multivariate photonics could enable and those which it cannot.

  \subsection{Scalability}
    One source of scaling limits stems from photonic weight bank hardware. The number of dimensions is limited by resonator finesse and a penalty related to the ability to weight neighboring channels independently, derived in~\cite{Tait:16scale} and determined to be 1.2 in~\cite{Tait:18twopole}. $Nf$ is limited by resonator FSR and a distortion penalty measured in~\cite{Tait:17oi} to be 4.3.
    Ref.~\cite{Soltani:10} showed oxide-clad microdisks with FSR of 57 nm (7.1~THz) and $Q$ of 80k, resulting in finesse of 2900. This means that multivariate photonic circuits using these resonators and WDM could scale to $N < 1500$ and $Nf < 1.65$~THz. There are other forms of multiplexing besides WDM, such as mode- and polarization-division multiplexing, that would further increase both of these limits.

    Scaling limits could also stem from algorithms. The model of projection moment vs. projection angle, Eq.~\eqref{eq:twoPetal}, can be straightforwardly extended to higher dimension by introducing additional angle variables; however, it is unclear if the model-based fitting method will also extend to higher dimensions in the presence of noise. For each additional dimension, two model parameters and thus two additional measurements are required. The fit quality can be improved by taking more than the minimum number of measurements or by averaging for longer per measurement. Further work might be able to determine if there is a limit to this approach or to develop other ways to exploit the information derived from measurements of statistical moments.

    There is also a question of how much need there is for a large number of antennas in actual applications. Today's WiMAX systems use less than 10 antennas, so there is not an immediate demand for information processors suited to thousands of inputs. Demands for bandwidth and dimensionality scaling will stem from future application areas still under study. mm-wave systems are a promising candidate application. In mm-wave systems, the center frequency is an order of magnitude higher than today's commercial 4G (60~GHz vs. 2.4~GHz or 5.0~GHz) meaning that spectrum bandwidths of interest could be in the 10s of GHz~\cite{Rappaport:11}. Correspondingly, the electromagnetic wavelength is an order of magnitude shorter, meaning there is finer spatial variation in the electromagnetic field that can only be resolved by denser antenna arrays. In the theoretical limit where antennas are spaced by $\lambda/2$, a 1 meter by 1 meter antenna array at 2.4~GHz would have no use for more than 16 x 16 = 256 antennas, whereas the same size array at 60~GHz could consist of 400 x 400 = 160,000 antennas before spatial information would become fully redundant. Using all of this information would require systems of many antennas backed by information processors that do not even remotely exist today.

  \subsection{Power}
    Power use is a central performance metric for handheld radios, vehicular radios, and ad-hoc mobile base stations employed, for example, by first responders. Multivariate photonic and DSP front-ends exhibit fundamentally different power scaling relationships that favor photonics when the number of channels and operating frequency are high.

    For a digital electronic front-end, every sample from each of $N$ antennas must be digitized before processing. This means the ADC power in DSP is $P_{dsp} = NfE_{adc}$, where $E_{adc}$ is the energy per ADC conversion. For a photonic front-end, only one ADC is required, and a laser is needed for every channel operating at a different wavelength. The expression for power is $P_{pho} = N P_{las} + f E_{adc}$, where $P_{las}$ is laser power. There is no dominant $Nf$ term. Power associated with photonic weighted addition does not scale with frequency because weighting occurs in passive filters and summation occurs through total power detection.

    The power scaling trends indicate that there is always an operating point in terms of $Nf$ above which photonics will use less power -- the essential question is whether that crossover point occurs low enough to be relevant to actual RF problems, such as mm-wave systems. This question demands quantification of those values across different systems and operating points.

  \subsection{Further experiments}
    The demonstration in this paper has limitations that could be addressed by future work to strengthen the claims. Firstly, only blind PCA was shown. While this is a step towards the more useful independent component analysis, a direct demonstration of BSS would carry considerably more significance. Further work on any blind multivariate capabilities, such as BSS, could make use of the methods introduced here for constraining observability through synchronization (Sec.~\ref{sec:blindEnvironment}).

    Secondly, bandwidth here was limited by impedance mismatches in the electrical subsystem (i.e. PPG, splitter, delay, RF amplifiers, modulators), which cause reflections resulting in signal distortion, cross-talk, and spurious correlation. These effects are exacerbated at higher bandwidths since the frequency response of discrete components is not consistent device to device. Monolithic electrical integration would have a significant impact on this limitation.

    Thirdly, signal-to-noise values here were relatively poor, which is attributable to the high loss in the RF photonic link. As seen in Table~\ref{tab:results} and Fig.~\ref{fig:metaData}, noise was a significant contributor to error. The dominant contribution to loss comes from non-optimized fiber-to-chip coupling (estimated 15~dB per facet). Fiber-to-chip loss could be eliminated by monolithic photonic integration where modulators and detectors are co-integrated with the weight banks.

\section{Conclusion}
  We introduce concepts, key methods, and preliminary results in multivariate photonics: the application of multivariate statistical analysis to analog photonic processing. As a demonstration, we fabricated a silicon photonic circuit that identifies the principal components of two-dimensional, 1.0~GHz signals with less than 14\% error. The analog photonic domain presents performance characteristics beyond the foreseeable state-of-the-art in digital signal processing, particularly for multi-dimensional (e.g. multi-antenna) tasks.

  Dimensionality-reduction underlies blind source separation, a widely desired capability in wireless systems that presents a significant information processing challenge. Moving dimensionality-reduction to the analog domain can break harsh performance tradeoffs associated with analog-to-digital conversion; however, it also constrains the type of information that can be observed.

  A new type of algorithm based on measurements of statistical moments was developed to work within realistic constraints on waveform observability. It was evaluated on photonic hardware using novel experimental techniques to enforce realistic conditions of observability in a lab. The concepts and methods demonstrated lay a groundwork for experimental research into multivariate photonic algorithms, hardware, and applications.

\section*{Acknowledgment}
  This work supported by the National Science Foundation (NSF) Enhancing Access to the Radio Spectrum (EARS) program, Grant No. (ECCS 1642962). Fabrication support was provided via the Natural Sciences and Engineering Research Council of Canada (NSERC) Silicon Electronic-Photonic Integrated Circuits (SiEPIC) Program. Devices were fabricated by Cameron Horvath and Mirwais Aktary at Applied Nanotools, Inc., Alberta, Canada.

\bibliography{multivariatePhotonics}

\end{document}